\documentclass[article]{aa}
\usepackage{graphics,latexsym,epsfig,graphicx, txfonts}     
\usepackage{subfigure}
\begin{document}
\title{A simple analysis of halo density profiles using gravitational lensing time delays} 
\author{Benjamin M. Dobke \& Lindsay J. King} 
\institute{Institute of Astronomy, University of Cambridge, Madingley Rd, Cambridge CB3 0HA\\ }
\date{Received ... / Accepted ...} 
\authorrunning{Dobke \& King} 
\titlerunning{Halo Density Profiles from Time Delays} 
\abstract{Gravitational lensing time delays depend upon the Hubble constant and the density distribution of the lensing galaxies.  This allows one to either model the lens and estimate the Hubble constant, or to use a prior on the Hubble constant from other studies and investigate what the preferred density distribution is. Some studies have required compact dark matter halos (constant M/L ratio) in order to reconcile gravitational lenses with the HST/WMAP value of the Hubble constant (72 $\pm$ 8 km\,s$^{-1}$ Mpc$^{-1}$ and 72 $\pm$ 5 km\,s$^{-1}$ Mpc$^{-1}$, respectively).  This is in direct contradiction with X-ray, stellar dynamical, and weak lensing studies, which all point towards extended halos and isothermal density profiles.  In this work, we examine an up-to-date sample of 13 lensing galaxies resulting in a data set consisting of 21 time delays.  We select systems in which there is a single primary lensing galaxy (e.g. excluding systems undergoing mergers). Analysis is performed using analytic models based upon a power-law density profile ($\rho\propto r^{-\eta}$) of which the isothermal profile is a special case ($\eta$ = 2). This yields a value of $\eta$ = 2.11$\pm0.12$ (3$\sigma$) for the mean profile when modeling with a prior on the Hubble constant, which is only consistent with isothermality within 3 $\sigma$.  Note that this is a formal error from our calculations, and does not include the impact of sample selection or simplifications in the lens modeling. We conclude that time delays are a useful probe of density profiles, in particular as a function of the environment in which the lens resides, when combined with a prior on the Hubble constant. 
 
} 
\def\A{{\cal A}}
\def\eck#1{\left\lbrack #1 \right\rbrack}
\def\eckk#1{\bigl[ #1 \bigr]}
\def\rund#1{\left( #1 \right)}
\def\abs#1{\left\vert #1 \right\vert}
\def\wave#1{\left\lbrace #1 \right\rbrace}
\def\ave#1{\left\langle #1 \right\rangle}
\def\arcsecf {\hbox{$.\!\!^{\prime\prime}$}}
\def\arcminf {\hbox{$.\!\!^{\prime}$}}
\def\bet#1{\left\vert #1 \right\vert}
\def\vp{\varphi}
\def\vt{{\vartheta}}
\def\map{{$M_{\rm ap}$}}
\def\d{{\rm d}}
\def\mj{$\rm {m_{j}}$}
\def\mk{$\rm {m_{k}}$}
\def\col{$\rm {m_{j}}-\rm {m_{k}}$}\def\eps{{\epsilon}}
\def\vc{\vec} 
\def\s{{\rm d}}
\def\s{{\rm s}}
\def\t{{\rm t}}
\def\E{{\rm E}}
\def\L{{\cal L}}
\def\i{{\rm i}}
\def\seps{{\sigma_{\epsilon}}}
{\catcode`\@=11
\gdef\SchlangeUnter#1#2{\lower2pt\vbox{\baselineskip 0pt \lineskip0pt
  \ialign{$\m@th#1\hfil##\hfil$\crcr#2\crcr\sim\crcr}}}}
\def\gtrsim{\mathrel{\mathpalette\SchlangeUnter>}}
\def\lesssim{\mathrel{\mathpalette\SchlangeUnter<}}      
\maketitle
\section{Introduction}

In 1964, Refsdal first suggested that time delays between gravitationally lensed images could be used to determine the Hubble constant $H_{0}$.  The appeal of this method was the fact that it did not depend upon the local distance scale and was a relatively straightforward single-step process.  In addition, the time delay theory upon which it is based was derived from the established theory of general relativity.  Only in recent years, however, with the availability of high resolution radio and optical imaging, has the method begun to yield results.

When focusing on individual lensing galaxies, a large number of lens modeling studies (e.g. Impey et al. 1998; Williams \& Saha 2000; Winn et al. 2002) have obtained values of $H_{0}$ that have been lower than that from the \textit{Hubble Space Telescope (HST) Key Project} (Freedman et al. 2001) which yields a value of $H_{0}$ = 72 $\pm$ 8 ${\rm km}\,{\rm s}^{-1}{\rm  Mpc}^{-1}$, or indeed the \textit{Wilkinson Microwave Anisotrophy Probe (WMAP)} which yields a very similar $H_{0}$ = 72 $\pm$ 5 ${\rm km}\,{\rm s}^{-1}{\rm Mpc}^{-1}$, although this assumes a flat universe (Spergel et al. 2003).  In particular, a few studies have inferred that if one assumes that lenses have an isothermal mass distribution (flat rotation curves) the derived value is $H_{0}\sim 50{\rm km}\,{\rm s}^{-1}{\rm Mpc}^{-1}$, whereas if they assume a constant mass-to-light ratio (M/L), which results in a falling rotation curve, then $H_{0}\sim 70 {\rm km}\,{\rm s}^{-1} {\rm Mpc}^{-1}$, which is in agreement with the \textit{HST/WMAP} finding (Kochanek 2002; others e.g. Impey et al. 1998; Rusin et al. 2003).  This constant M/L ratio implies a compact dark matter halo.  

However, other studies such as weak lensing (e.g. Guzik \& Seljak 2002), stellar dynamics (Rix et al. 1997; Romanowsky \& Kochanek 1999; Gerhard et al. 2001; Treu \& Koopmans 2002), and X-ray galaxy measurements (e.g. Fabbiano 1999; Lowenstein \& White 1999), are consistent with near isothermal density profiles (i.e. extended dark matter halos).  Note however that in the strong lensing regime we are probing the inner few kpc of galaxies (i.e. within the Einstein radius), whereas other studies, such as galaxy-galaxy lensing, are sensitive to more extended scales.

The aim of this paper is to study whether time delays from multiple image lens systems statistically point towards extended dark matter halos which are inferred from other measurements, when an external prior is adopted for $H_{0}$.   We perform the study without the complexity of detailed lens modeling in an effort to provide insight with a simple model.  The outline of this paper is as follows. In \S2 we introduce two analytic models for the gravitational time delay based upon an SIS potential and general power law potential.  \S3 will discuss the effect of ellipticity and shear on the time delays.  In \S4 we discuss the lens sample used as the data set in the investigation, and in \S5 we explain the analysis of this data set and present results. \S6 draws some conclusions based upon the findings.

\section{Analytic Models for the Gravitational Time Delay}

The light-travel time for an image of a gravitational lens system, is given by (see e.g. Narayan \& Bartelmann 1997)

\begin{equation}
t(\vec{\theta}) = \bigg[\frac{1+z_{l}}{c}\bigg]\bigg[\frac{D_{l} D_{s}}{D_{ls}}\bigg]
\bigg[
\frac{1}{2} \left|\vec{\theta}- \vec{\beta}\right|^{2} - \psi(\vec{\theta})\bigg]
\label{equ:time} 
\end{equation}

\noindent where $z_{l}$ is the redshift of the lensing galaxy, $D_{l}$ and $D_{s}$ are the angular diameter distances to the lens and source respectively, and $D_{ls}$ is the angular diameter distance between the lens and source.  The two dimensional vectors $\vec{\theta}$ and $\vec{\beta}$, are the (angular) image and source positions respectively and the effective potential at the locations of images is $\psi(\vec{\theta})$. Note that the light-travel time has a geometrical component and a component due to gravitational time dilation (Shapiro delay) which depends on the gravitational potential. Since angular diameter distance scales as $c/H_{0}$, where $H_{0}$ is the Hubble parameter at the present epoch, it follows that $t\propto 1/H_{0}$. The time delay between two images, say A and B, is the difference in their light-travel time i.e. $\Delta t = t_{A} - t_{B}$.  In order to calculate this quantity we must adopt a specific effective potential.

\subsection{Time Delay from an SIS Potential\label{gen}}
\begin{figure}[t!]
\begin{center}
\epsfig{file=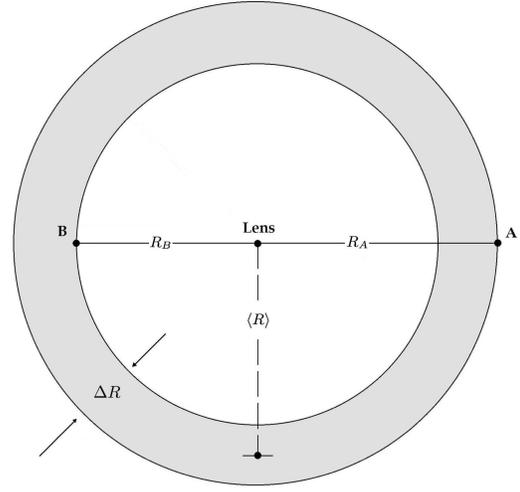,scale=.4}
\caption{Diagram of a time delay lens with two images, A and B.  The lensing galaxy is at the centre with the images at radii $\left|\vec{\theta}\right|$ which we denote by $R_{A}$ and $R_{B}$.  The images bound an annulus of width  $\Delta R = R_{A} - R_{B}$, with average radius $\langle R \rangle = ( R_{A} + R_{B})/2$ (as in Kochanek \& Schechter 2004).}
\label{plot1}
\end{center}
\end{figure}

  The singular isothermal sphere (SIS) is a simple yet surprisingly realistic starting point for modeling lenses. 
Its lensing potential has the form:

\begin{equation}
\psi(\vec{\theta}) = b\left|\theta\right|
\label{equ:sis}
\end{equation}

\noindent where

\begin{equation}
b = 4\pi\frac{D_{ls}\sigma^{2}}{D_{s} c^{2}}
\label{equ:lensstrength}
\end{equation}

\noindent is a deflection scale determined by the geometry.  For a $|\beta| < b$, the SIS produces two collinear images at radii $R_{A}$ = $|\beta| + b$ and $R_{B}$ =  $b - |\beta|$ on opposite sides of the lens galaxy, as in Fig.\,\ref{plot1}. The time delay between these two images can be shown to be

%


%



\begin{equation}
\Delta t_{SIS} = \frac{1}{2}\, \bigg[\frac{1+z_{l}}{c}\bigg]\bigg[\frac{D_{l} D_{s}}{D_{ls}}\bigg](R_{A}^2 - R_{B}^{2})
\label{equ:timedelaysis}
\end{equation}

\noindent which  depends upon the annulus bounded by the two image positions, as in Fig.\,\ref{plot1}, and as derived in Witt et al. (2000).

\subsection{Time Delay from a General Power Law Potential\label{spec}}
A more general approach is to use an effective potential that allows for different slopes ($\eta$) for the radial density profile
\begin{equation}
\psi (\vec{\theta}) = \frac{b^{2}}{(3-\eta)}\bigg(\frac{\left|\theta\right|}{b}\bigg)^{3-\eta}
\label{equ:pl}
\end{equation}
of which the SIS is a special case ($\eta=2$), with more centrally concentrated mass distributions having higher values of $\eta$ (tending to a point mass with $\eta=3$). The deflection scale $b$ is given by (e.g. the contribution of Kochanek in Kochanek, Schneider \& Wambsganss 2004)
\begin{equation}
b=\left[\frac{R_{A}+R_{B}}{R_{A}^{2-\eta}+R_{B}^{2-\eta}}\right]^\frac{1}{\eta-1},
\end{equation}
and the convergence is
\begin{equation}
\kappa(\theta) =\frac{3-\eta}{2} \left(\frac{\left|\vec{\theta}\right|}{b}\right)^{1-\eta}.  
\end{equation}
Previous studies have found that lens galaxies are well described by such power-law profiles, in which $\rho\propto r^{-\eta}$ (e.g. Witt et al. 2000; Rusin et al. 2003).
To obtain a time delay from the power-law lens potential, we use the expression detailed in Kockanek (2002):  
\begin{eqnarray}
&&\Delta t_{AB} =   \bigg[\frac{1+z_{l}}{c}\bigg]\bigg[\frac{D_{l} D_{s}}{D_{ls}}\bigg]
\nonumber\\\nonumber&&\times\Bigg\{(1-\langle \kappa \rangle)\bigg[-\frac{1}{2}(R_{A}^{2} - R_{B}^{2}) + R_{A}R_{B}\ln\frac{R_{A}}{R_{B}}\bigg]\\&&- 2\int_{R_{A}}^{R_{B}}udu[\kappa(u) - \langle \kappa \rangle]\ln\frac{u}{R_{B}}\Bigg\}
\label{equ:timedelayab}
\end{eqnarray}
\noindent where

\begin{equation}
\langle \kappa \rangle = \frac{2\int_{R_{A}}^{R_{B}} \kappa(u)udu}{R_{B}^{2}-R_{A}^{2}}
\label{equ:kappa}
\end{equation}

\noindent is the mean surface mass density in the annulus bounded by images A and B, in units of the critical surface mass density.

In the analysis that follows in \S5, we use $\Omega_{M}$ = 0.3, $\Omega_{\Lambda}$ = 0.7, $\Omega_{K}$ = 0, which enter through the angular diameter distances. 

\section{The effect of Galaxy Ellipticity and External Shear}

In the preceding section we introduced a spherically symmetric effective potential which will be used to model the lensing galaxy of all systems in the data set, including four-image systems. Our goal is to obtain an estimate of the average slope of the density profiles using a $simple$ approach, comparing systems in a homogenous manner and without detailed modeling of ellipticity or external shear.  We will now, however, briefly discuss the effect these have on the corresponding time delays using Fermat surfaces for illustration.

\begin{figure}
\begin{center}
\epsfig{file=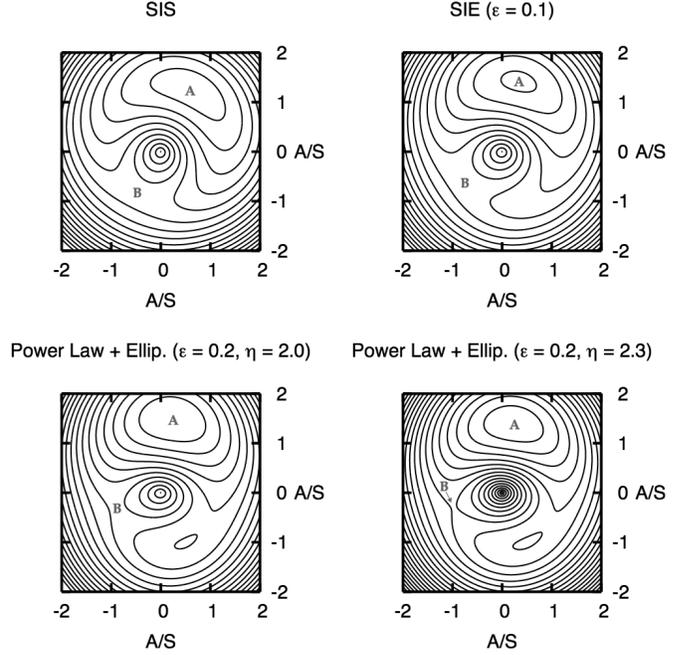,scale=0.51}
\caption{A plot of four Fermat time delay surfaces for varying effective potentials.  The images are to be found at the stationary points of the surface, and are displayed as A, B, and C.  The corresponding time delays in units of days are as follows; (a) $\Delta t_{SIS}$ = 51.2; (b) $\Delta t_{SIE, \epsilon=0.1}$ = 57.1; (c) $\Delta t_{\eta=2.0,\epsilon=0.2}$ = 67.9 (AB), $\Delta t_{\eta=2.0,\epsilon=0.2}$ = 48.8 (AC); (d) $\Delta t_{\eta=2.3, \epsilon=0.2}$ = 67.2 (AB), $\Delta t_{\eta=2.3, \epsilon=0.2}$ = 46.8 (AC)}
\label{plot4}
\end{center}
\end{figure}

Fig.\,\ref{plot4} shows four Fermat surfaces, as derived from equation (\ref{equ:time}), each of which displays a different image configuration dependent on the effective potential and parameters used.  For all four plots the source position, velocity dispersion, lens/source redshifts and Hubble constant were kept fixed.  We consider the changes in time delays  of the images A, B and C, present at the stationary points of the surface (note that there may be other stationary points).  We investigated values of ellipticity up to 0.2, this being a moderate ellipticity for a lens galaxy.     


Fig.\,\ref{plot4}a begins with a standard SIS potential, with a time delay of $\Delta t_{AB}$ = 51.2 days.  An ellipticity is then introduced into the effective potential (an SIE with $\epsilon$ = 0.1), as seen in (b), giving a time delay of $\Delta t_{AB}$ = 57.1 days.  Finally, in plots (c) and (d), we consider the impact of changing the slope of the effective potential. Keeping $\epsilon$ = 0.2, the slope of the density profile is allowed to change from $\eta$ = 2.0 to $\eta$ = 2.3 (corresponding to a slope of 1.0 to 0.7 in the effective potential).  The time delay is $\Delta t_{AB}$ = 67.9 days and $\Delta t_{AC}$ = 48.8 days for the $\eta$ = 2.0 case, and $\Delta t_{AB}$ = 67.2 days and $\Delta t_{AC}$ = 46.8 days, for the $\eta$ = 2.3 case. 
In the limit $\eta$ tends to 3.0, $\Delta t_{AB}$ = 66.7 days and $\Delta t_{AC}$ = 44.5 days .

From this we highlight two key features:  firstly, the time delays between images A and B  change by less than 10\% when comparing the SIS to the SIE with a small ellipticity.  Secondly, when we consider the change in the time delays as $\eta$ goes from 2.0 to 2.3, keeping $\epsilon$ constant, we observe very little change at all.  We pick these particular values for illustrative purposes only. Indeed, moving beyond 2.3 to higher values of $\eta$ had a similarly small effect. 


As noted by Witt et al. (2000) for a generalised isothermal lens (allowing for general angular structure irrespective of ellipticity) the Hubble constant is simply related to the image positions and time delay along with the source and lens redshifts, not requiring detailed models. 
When the profile deviates from isothermal, and ellipticity is included, detailed modeling cannot be avoided. However, for images which are not close pairs, they derived scaling relationships which show that for moderate ellipticities, the time delay is well approximated by the isothermal delay modified by a factor which depends on the slope of the density profile or effective potential.  The scaling relations themselves follow the form ($\eta$-1) for images at $\sim 180^{\rm o}$, and ($\eta$-1)/(3-$\eta$) for those at $\sim 90^{\rm o}$, and are multiplied with the isothermal time delay $\Delta t_{sis}$, of equation (\ref{equ:timedelaysis}). 

To estimate the effect of these scalings we generated image pairs (at both $\sim 90^{\rm o}$ and $\sim 180^{\rm o}$) from Fermat surfaces derived from the parameters $\epsilon$ = 0.2 and $\eta$ = 2.3 (see Fig. \ref{plot4}d).  Fixing $H_{0}$ = 72 km s$^{-1}$ Mpc$^{-1}$, we recovered $\eta$ from the time delay expression in \S2.2 (eqn. \ref{equ:timedelayab}) for both sets of image pairs.  We found that when using the $180^{\rm o}$ scaling we recovered a value of $\eta$ = 2.32 for image pair AC, and $\eta$ = 2.46 for AB.  Using the $\sim 90^{\rm o}$ scaling yielded $\eta$ = 2.18 for AB.  
In the limit  $\eta$ tends to 3.0, the maximum error in recovered $\eta$ is less than 20\%; note that using equation (\ref{equ:timedelayab}) as we do below, rather than taking into account $90^{o}$ scaling is actually more accurate in recovering the slope.

When considering the effect of the $\sim 90^{\rm o}$ scaling on the real image data of Table 1, we found that in general the scaling gave values of $\eta$ both higher and lower than those derived from the $\sim 180^{\rm o}$ scaling.  Moreover, the difference in the derived values for $\eta$ from both scalings was less than 7\%. 

From the above, we see that while using a spherically symmetric effective potential or mass profile in the analysis of four image systems is certainly to be considered an approximation, it would appear not to be grossly inaccurate when considering the variation of derived parameters.

External shear also impacts on the analysis, as discussed by Witt et al. (2000); they show that if there is no knowledge of the angle between the shear and the lensing galaxy's major axis, then a ~10\% shear would lead to a ~10\% uncertainty in time delay.  However, strong external shear would lead to more extreme uncertainties, which will lead to the exclusion of certain systems from the working data sample.

\begin{table*}

\begin{center}
\title{TABLE 1\\ \scriptsize{Observational Data for Time Delay Lenses} \\ $ $\\}
{\scriptsize
\begin{tabular}{lccr@{ }lccc}\hline\hline
&\\
Lens/Components
& $z_{l}$ 
& $z_{s}$ 
& \multicolumn{2}{c}{$\Delta t_{i,j}$ (days)}  
& $R_{j}$ (arcsec)
& $R_{i}$(arcsec)
& References \\
\hline
&&&&&& \\
PG 1115+080 (A-B) & 0.31 & 1.72 & 11.7&$\pm$ 1.2 & 1.147 & 0.950 & 1 \\ 
PG 1115+080 (B-C) & 0.31 & 1.72 & 25.0& $\pm$ 1.6 & 0.950 & 1.397 & 1 \\
PG 1115+080 (A-C) & 0.31 & 1.72 & 13.3&$\pm$ 1.0 & 1.147 & 1.397 & 1 \\
HE 2149-2745 & 0.50 & 2.03 & 103 &$\pm$ 12 & 0.344 & 1.354 & 2 \\
HE 1104-1805 & 0.73 & 2.32 & 161 &$\pm$ 7 & 1.099 & 2.095 & 3 \\
SBS 1520+530 & 0.72 & 1.86 & 130 &$\pm$ 3 & 0.385 & 1.207 & 4 \\
Q0957+561 & 0.36 & 1.41 & 417 &$\pm$ 3 & 1.036 & 5.220 & 5 \\
B1422+231 (A-B) & 0.34 & 3.62 & 1.5 &$\pm$ 1.4 & 1.014 & 0.961 & 6 \\
B1422+231 (B-C) & 0.34 & 3.62 & 8.2 &$\pm$ 2.0 &  0.961 & 1.056 & 6 \\
B1422+231 (A-C) & 0.34 & 3.62 & 7.6 &$\pm$ 2.5 & 1.014 & 1.056 & 6 \\
B1600+434 & 0.42 & 1.59 & 51 &$\pm$ 4 & 0.250 & 1.140 & 7 \\
RXJ 0911+0551 & 0.77 & 2.80 & 146 &$\pm$ 4 & 0.866 & 1.327 & 8 \\
B0218+357 & 0.68 & 0.96 & 10.5 &$\pm$ 0.2 & 0.100 & 0.240 & 9 \\
PKS 1830-211 & 0.89 & 2.51 & 26 &$\pm$ 4 & 0.320 & 0.670 & 10 \\
B1608+656 (B-A) & 0.63 & 1.39 & 76 &$\pm$ 10 & 1.091 & 1.085 & 11 \\
B1608+656 (B-C) & 0.63 & 1.39 & 30 &$\pm$ 7 & 1.091 & 0.682 & 11 \\
B1608+656 (B-D) & 0.63 & 1.39 & 36 &$\pm$ 7 & 1.091& 1.451 & 11 \\
HE 0435-1223 (A-B) & 0.45 & 1.69 & 8.0 &$\pm$ 0.8 & 1.298 & 1.168 & 12 \\
HE 0435-1223 (A-C) & 0.45 & 1.69 & 2.1 &$\pm$ 0.8 & 1.298 & 1.065 & 12 \\
HE 0435-1223 (A-D) & 0.45 & 1.69 & 14.4 &$\pm$ 0.8 & 1.298 & 1.302 & 12 \\
FBQ 0951+2635 & 0.24 & 1.25 & 16.0 &$\pm$ 2.0 & 0.221 & 0.879 & 13 \\
\hline 

\end{tabular}
} 
\scriptsize{
\begin{verse}
\begin{center}

References. - (1) Barkana 1997; (2) Burud et al. 2002a; (3) Ofek \& Maoz 2003; (4) Burud et al. 2002b; \\ (5) Kundi\'{c} et al. 1997; (6) Patnaik \& Narasimha 2001; (7) Burud et al. 2000; (8) Hjorth et al. 2002; \\(9) Biggs et al. 1999; (10) Lovell et al. 1998; (11) Fassnacht et al. 2002; (12) Kochanek et al. 2006; \\ (13) Jakobsson et al. 2005
\end{center}
\end{verse}
}
\end{center}

\end{table*}

\section{Data: The Lensing Sample}

To date, there are 13 gravitational lensing systems with well measured time delays.  However, a number of these have more than two bright images. In total, there are 21 time delay measurements, and it is these that constitute the initial data set for our investigation (see Table 1).  The primary source of errors in the data originate from the time delay; most of the errors in image positions are on the order of a few percent, and as such are negligible in comparison.

In the analysis that follows in \S5, there are three lens systems that have been excluded from the working data set.  These are RXJ0911+0551, Q0957+561 and B1608+656.  For RXJ0911+0551, the main lensing galaxy has a smaller, less massive galaxy inside the Einstein ring, and is also embedded in a large X-ray cluster (Morgan et al. 2001).  Q0957+561 is a special case as the lensing galaxy is a BCG (brightest cluster galaxy). In both cases, there are $strong$ external perturbations that would require a much more complex effective potential to successfully attempt to model the time delay, and as such would not be well modeled by the expressions in \S2.  Weaker external shear results in far less uncertainty in the modeled time delays, as discussed in \S3.  An additional system that has notable complexity is B1608+656.  Although simple lensing potentials appear to produce an acceptable fit (Fassnacht et al. 2002) we exclude it from the sample considered since the lensing galaxy appears to be undergoing a merger with a smaller galaxy.  In addition to the above, the time delays PG1115+080 (B-C) and B1422+231(A-C) have been excluded since there is redundancy with the other delays quoted for the particular system. 
While each lens system obviously has only one true value for the slope of its density profile, each delay that we use from that system is an independent probe of that slope, with corresponding error. 

Finally, we are left with 10 systems and a total of 14 time delays for use in the analysis.  Note that the systems in the lens sample of the Kochanek (2002) semi-analytic study are PG1115+080, SBS1520-530, B1600+434, and HE2149-2745, and this sub-sample will also be examined in the next section.

\section{Analysis and Results}

The analysis of the data sample was performed with equation (\ref{equ:timedelayab}) of \S2.  For this simple analysis we exclude the effects of ellipticity and shear which allows us to use the time delay expression (\ref{equ:timedelayab}).  As mentioned in \S3, inclusion of either would require detailed modeling of each system individually defeating the purpose of a simplified approach. For a particular lens system, we use a single component lens model that accounts for the sum of the dark and luminous components, and determine the time delay for particular values of $H_{0}$ and $\eta$ given the image positions along with source and lens redshifts.  These values can then be compared to the observed delay using the $\chi^{2}$ statistic as follows:

\begin{equation}
\chi^{2} = \frac{(\Delta t_{i,j, Mod} - \Delta t_{i,j, Obs})^{2}}{\sigma_{Obs}^{2}}
\label{equ:chi}
\end{equation}

\noindent where $\Delta t_{i,j, Mod}$ is the modeled time delay from equation (\ref{equ:timedelayab}), $\Delta t_{i,j, Obs}$ is the observed time delay for a particular system, and $\sigma_{Obs}^{2}$ is the error in that observed time delay.  This process allows us to build up a grid of $\chi^{2}$ values in a 2-D plane of $H_{0}$ and $\eta$. As an additional constraint, we add in a gaussian prior to expression (\ref{equ:chi}) based on the HST Key Project value $H_{0}$ = 72 $\pm$ 8 km s$^{-1}$ Mpc$^{-1}$:

\begin{equation}
\chi^{2}_{prior} = \frac{(72 - H_{0})^{2}}{8^{2}}
\label{equ:prior}
\end{equation}

\begin{figure}[t!]
\begin{center}
\epsfig{file=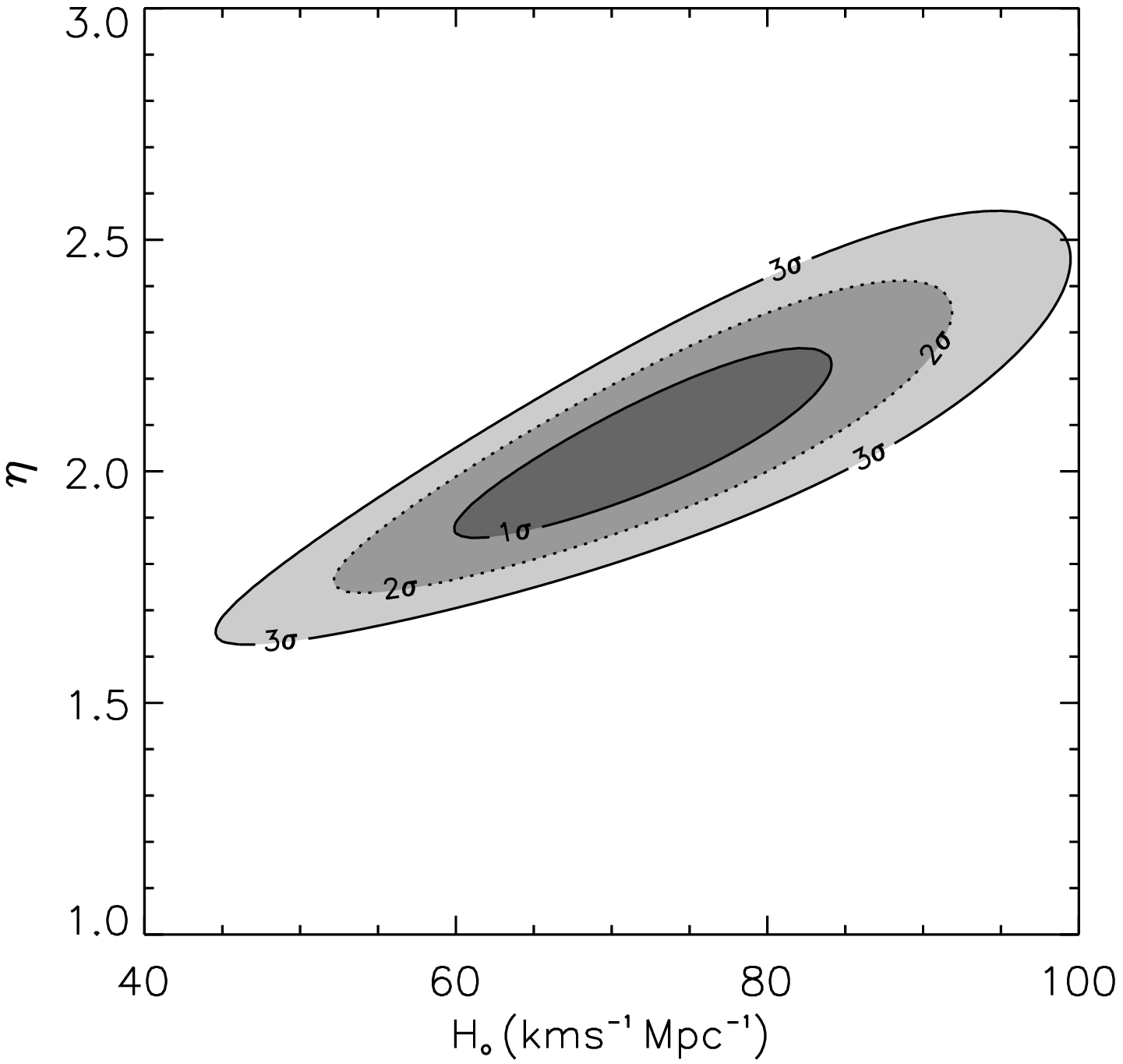,scale=0.61}
\caption{The $\Delta \chi^{2}$ limits for $H_{0}$ vs. $\eta$ for the time delay system PG1115+080 (A-B), including a prior on $H_{0}$ from the HST Key Project.  Note the clear degeneracy that exists between $H_{0}$ and $\eta$, where more centrally concentrated mass distributions (larger $\eta$) will result in time delays that produce a higher value for $H_{0}$ }
\label{plot2}
\end{center}
\end{figure}

\begin{table*}
\begin{center}
\title{TABLE 2\\ \scriptsize{Best Fit $\eta$ Parameter for Data Sample} \\ $ $\\}
{\scriptsize
\begin{tabular}{l|ccc}\hline\hline
&\\
Lens/Components
& Best Fit $\eta$ (inc. prior) 
& 1$\sigma$ error\\
\hline
&&& \\
PG 1115+080 (A-B) & 2.25 & +0.30/-0.27 \\ 
PG 1115+080 (A-C) & 1.92 & $\pm$ 0.18  \\
HE 2149-2745 & 2.56 & +0.44/-0.41  \\
HE 1104-1805 & 1.77 & $\pm$ 0.13  \\
SBS 1520+530 & 2.40 & +0.30/-0.25  \\
B1422+231 (A-B) & 1.63 & +0.92/-0.63 \\
B1422+231 (B-C) & 2.89 & +0.11/-0.34  \\
B1600+434  & 2.21 & +0.32/-0.24  \\
B0218+357 & 2.60 & $\pm$ 0.30  \\
PKS 1830-211 & 1.88 & +0.40/-0.28  \\
HE 0435-1223 (A-B) & 1.70 &  $\pm$ 0.12 \\
HE 0435-1223 (A-C) & 1.67 & $\pm$ 0.15  \\
HE 0435-1223 (A-D) & - *& - / - * \\
FBQ 0951+2635 & 2.27 & +0.43/-0.32 \\
\hline 
\end{tabular}
}
\scriptsize{
\begin{verse}
\begin{center}

* Contours failed to close in the range 1$<$ $\eta$ $<$ 3, yielding no error bounds.
\end{center}
\end{verse}
}
 
\end{center}
\end{table*} 

One can then relate these $\chi^{2}$ values to confidence levels based upon the number of degrees of freedom. Note that when dealing with a single time delay there is 1 data point ($N_{dp}$ = 1) and 2 free parameters ($N_{fp}$ = 2), and hence the number of degrees of freedom are -1 (since $N_{dof}$=$N_{dp}$-$N_{fp}$). This means that the problem is under-constrained in this instance. 


Despite the inclusion of the HST $H_{0}$ prior, there exists a notable degeneracy in the $H_{0}$ - $\eta$ plane and we highlight this in Fig.\,\ref{plot2} for the time delay PG1115+080 (A-B) .  
The direction of the degeneracy is such that the time delay data are consistent with a range of more (less) centrally concentrated mass distributions and higher (lower) values for $H_{0}$.  We include the best fit $\eta$ and corresponding 1$\sigma$ error for all the systems included in our sample (see Table 2).

\begin{figure}[t!]
\begin{center}
\epsfig{file=blank.eps,scale=0.605}
\caption{Enlarged plot of the combination of confidence limit (such as Fig.\,\ref{plot2}) for all 14 time delays, all of which include the HST $H_{0}$ value as a prior.  Plotted are the 68\%, 95\% and 99\% likelihood contours.  The centroid value corresponds to a density profile slope of $\eta$ = 2.11$\pm0.12$ (3$\sigma$) at $H_{0}$ =  71${\rm km}{\rm\,s}^{-1}{\rm  Mpc}^{-1}$.  Inset:  same plot but on the scale of Fig.\,\ref{plot2}.  While the degeneracy is still apparent, the extent of the contours is greatly reduced over that of Fig.\,\ref{plot2}, thus providing bounds on $\eta$.}
\label{plot3}
\end{center}
\end{figure}

\begin{figure}[h!]
\begin{picture}(100,-200)(100,-256)
\put(100,-120){\includegraphics[width=8.755cm]{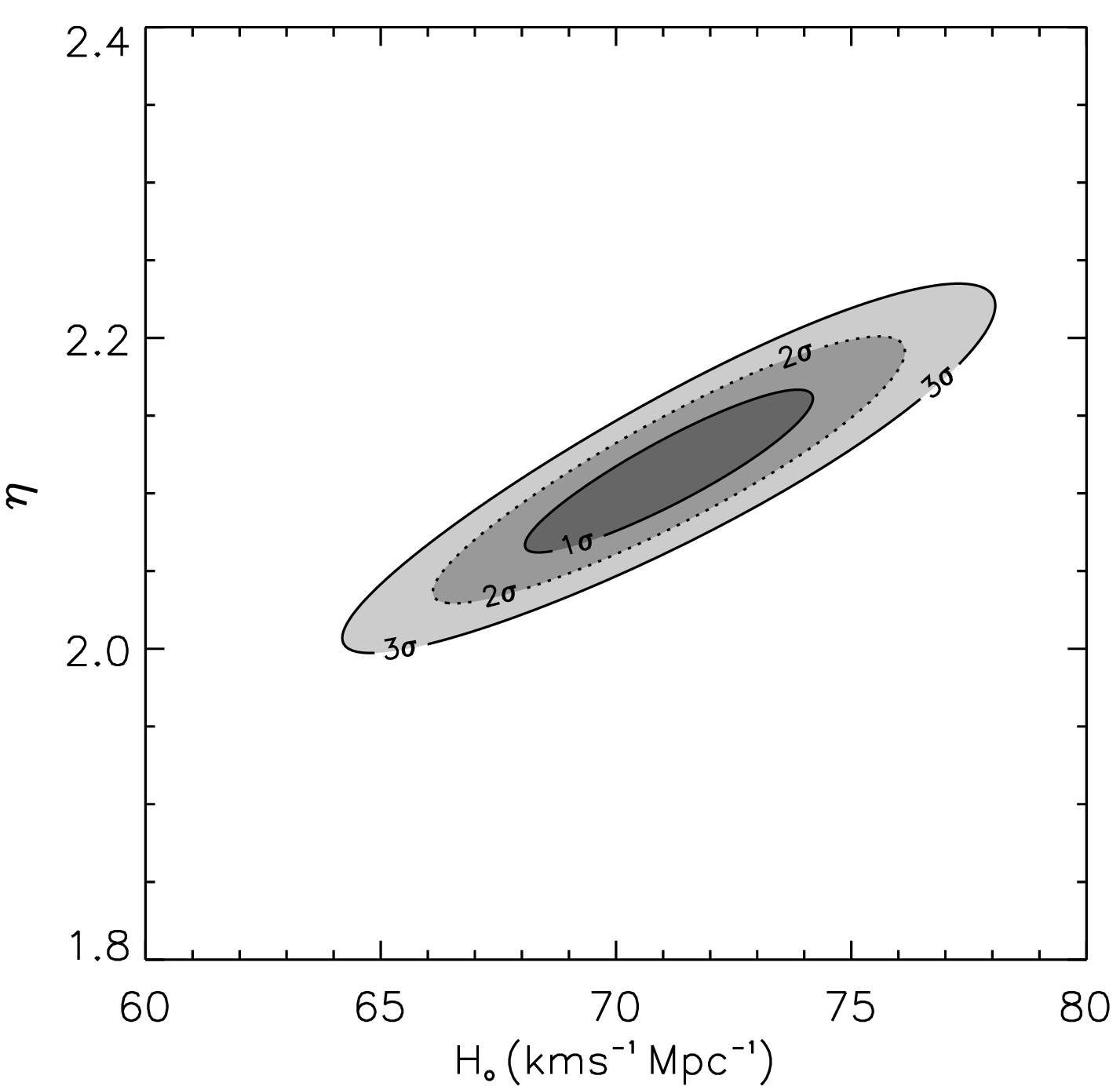}}
\put(140,25){\includegraphics[width=3.0cm]{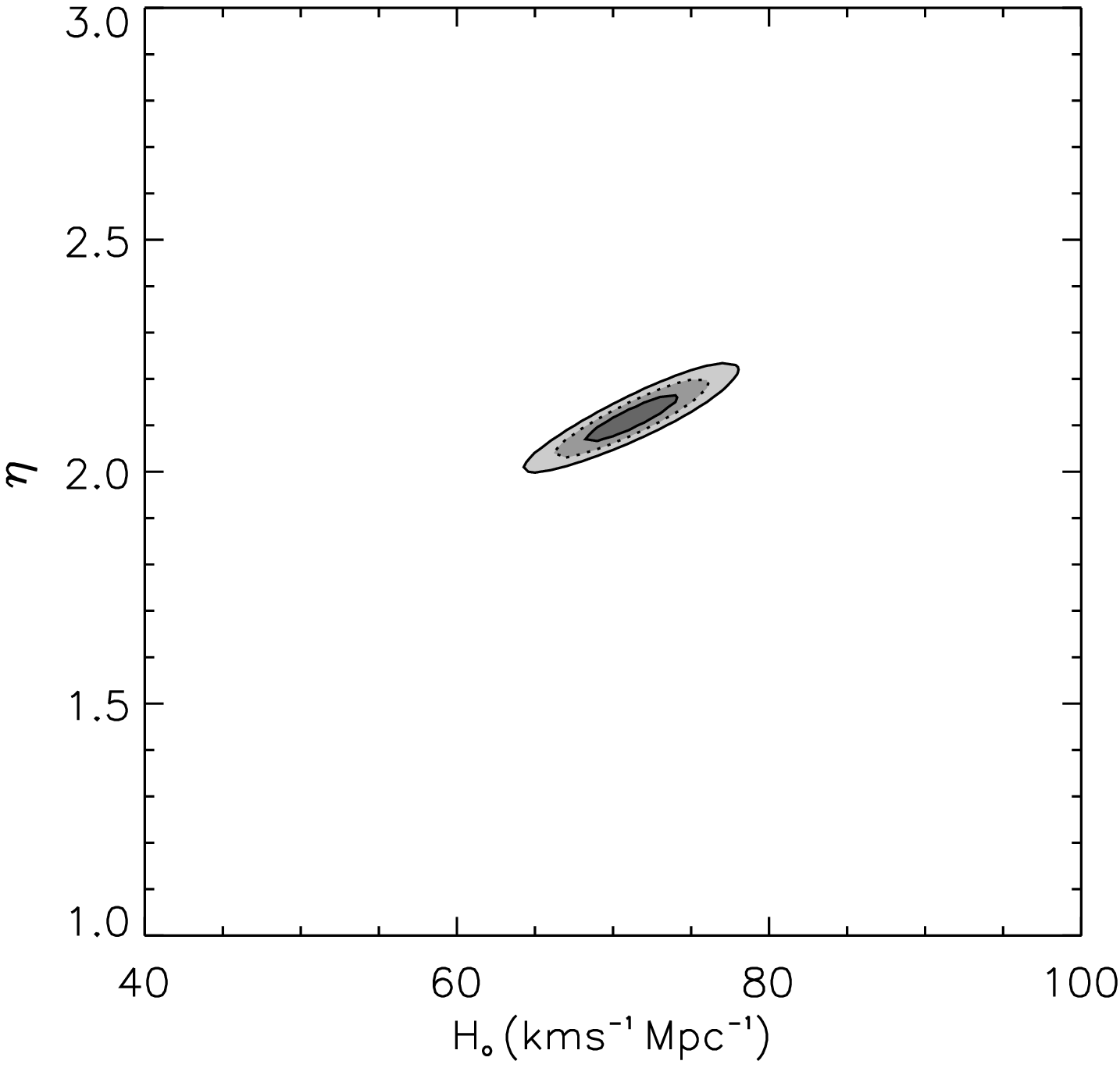}}
\end{picture}
\end{figure}

The result of combining the constraints from all 14 time delays, including the HST $H_{0}$ prior, is shown in Fig.\,\ref{plot3}.  Since we have a total of 14 data points and 2 free parameters,  there are 12 degrees of freedom .  The overall reduced $\chi^{2}$  for the data displayed in Fig.\,\ref{plot3} is $\chi^{2}_{\rm red} = 49.1$.  Such a high value for a combination of all the data reflects the simplified lens model, and the lack of a common density profile slope (i.e. that galaxies have a range of slopes for their profiles).  The extent of the contours is greatly reduced over that of Fig.\,\ref{plot2}, thus providing bounds on $\eta$. The best fit density profile slope is $\eta$ = 2.11$\pm0.12$ (3$\sigma$). If instead we 
exclude redundant time delays from our sample based on the largest fractional error rather than absolute error, we obtain the same result. This is also the case if we include every time delay, not excluding any which could be derived from the others.
The selection of our sample described in Section 4 above means that these results are relevant to 
lens systems in which there is a single primary lens - in other words to galaxies which are not undergoing mergers, or in very dense environments. Probing the density profiles of galaxies in complex environments requires detailed modeling to account for external perturbers. Note also that our error bars are formal errors from the calculations, and do not address the systematics of sample selection or simplifications in modeling. 

When the data are separated into time delays resulting from quad image systems and doubles, then from the 7 time delays coming from each we obtain $\eta$ = 1.79$\pm0.15$ (3$\sigma$) and $\eta$ = 2.23$\pm0.2$ (3$\sigma$) respectively. 

When considering just the four systems PG1115+080, SBS1520+530, B1600+434, and HE2149-2745, before application of the prior on the Hubble constant, we find a steeper best fit slope than for the remainder of the sample, in keeping with Kochanek (2002).
Kochanek et al. (2006) note that this may well be due to the environment of the lensing galaxies (e.g. halo stripping of group satellite galaxies).  Indeed, other authors have noted that lens galaxy environments could potentially have a great impact on halo profiles (Dalal \& Watson 2004; Keeton \& Zabludoff 2004). This also suggests that the profile slopes of the other systems we have excluded from our analysis due to the primary lens being in a very complex environment would be likely to increase the overall dispersion. We also note that if the 4 systems noted above are in addition excluded from the analysis then we obtain $\eta$ = 2.06$\pm0.14$ (3$\sigma$) including the prior on the Hubble constant. 


\section{Discussion and Conclusions}
Gravitational lens time delays directly depend upon the Hubble constant and the density distribution of the lensing galaxies.  This allows us to either model the lens and infer the Hubble constant, or to marginalize over the Hubble constant derived from other studies and determine the preferred density distribution. Our goal has been to estimate the typical slope of a massive galaxy's density profile, using simple models as in Witt et al. (2000).

Although some studies have indicated that it does not seem possible to reconcile an isothermal model for galaxy halos with the HST/WMAP value for the Hubble constant, rather requiring a compact halo (Impey et al. 1998; Kochanek 2002), our analysis indicates that, on average, this is not the case.  For the sample of relatively isolated galaxies which we consider, as defined in Section 4, the mean value was found to be $\eta$ = 2.11$\pm0.12$ (3$\sigma$), when incorporating a prior on the Hubble constant from the HST Key Project. The quoted error arises formally from the calculation outlined in Section 5 and does not include any bias due to sample selection, or due to the simplifications of lens modeling - for example, not fitting for ellipticity or external shear. 
Our finding is in agreement with a study of 22 lens galaxies using self-similar mass models, which resulted in $\eta$ = 2.07 $\pm$ 0.13 (1$\sigma$) (Rusin et al. 2003). The high reduced $\chi^{2}$ of our best fit to the data is a combination of the large amount of scatter in the density profile slopes of the sample, and over-simplistic models for the lenses. 
The steeper slope obtained for lenses resulting in double image configurations is perhaps surprising in that multiple imaging is more effective for steeper profiles; however, the cross-section for quads scales roughly as $\epsilon^{2}$, so this also a competing factor.
Previous studies (e.g. Witt et al. 2000; Kochanek 2002) have shown that two-image systems are relatively insensitive to quadrupole structure, and above we briefly considered how deviations from SIS impact on the time delay surface and estimates of the slope of the density profile.

We also note that the compact halo that tends to $\eta$ = 3 is not typical. However, when we consider the four systems PG1115+080, SBS1520-530, B1600+434, and HE2149-2745, there is a tendency towards a steeper than isothermal density profile, although in this case we are dealing with small number statistics with only four systems to constrain the $\eta$ parameter.  Kochanek et al. (2006) have recently pointed out that at least three of these four systems might have steeper density profiles as a consequence of environment - the distinction being whether at one extreme the galaxy is a satellite galaxy (and partially stripped of its dark matter halo) or at the other extreme is a central galaxy in a group. The analysis of Kochanek et al. (2006), coupled with our findings for the remaining time delay systems ($\eta$ $\sim$ 2), suggests that strong gravitational lensing time delays should be seen as placing valuable constraints on galaxy and group formation scenarios, reflecting the interplay between baryons and dark matter on these scales. One step to assess the impact of galaxy environment on time delays would be to investigate the lensing properties of fully simulated groups - are the profiles derived from simulated lensing time delays consistent with these? Such a study would allow us to address the impact of complex environments (e.g. for the systems we excluded from our sample) on galaxy profiles.

Ideally, more time delay measurements should be made to reduce the errors, and measurements made to increase the accuracy of the astrometry, particularly in systems where the lensing galaxy's position has notable uncertainties.  Moreover, increasing the sample size of time delay systems is crucial, as it puts further constraints on density profiles and reduces the random uncertainties due to large-scale structure. Looking towards the future, GAIA will uncover roughly fifty times more multiply imaged quasars than we have identified today (see Gaia study report); of these a few percent will have variable radio emission and hence be potential candidates for radio time delay systems.

\acknowledgements
We thank Wyn Evans, Antony Lewis, Peter Schneider and Steve Warren for discussions, and the anonymous referees for their constructive comments.  This work was supported by PPARC through a PhD studentship (BMD), and by the Royal Society (LJK). 

\def\ref#1{\bibitem[1998]{}#1}

\end{document}